# Investigating the Significance of Bellwether Effect to Improve Software Effort Estimation

Solomon Mensah[1], Jacky Keung[1], Stephen G. MacDonell[2], Michael F. Bosu[3] and Kwabena E. Bennin[1]
[1]*Department of Computer Science, City University of Hong Kong, Hong Kong, China*
[2]*Department of Information Science, University of Otago, Dunedin 9054, New Zealand*
[3]*Centre for Business, Information Technology and Enterprise, Wintec, Hamilton, New Zealand*
{smensah2-c, kebennin2-c}@my.cityu.edu.hk, Jacky.Keung@cityu.edu.hk,
stephen.macdonell@otago.ac.nz, michael.bosu@wintec.ac.nz

**Abstract**

*Bellwether effect refers to the existence of exemplary projects (called the Bellwether) within a historical dataset to be used for improved prediction performance. Recent studies have shown an implicit assumption of using recently completed projects (referred to as moving window) for improved prediction accuracy. In this paper, we investigate the Bellwether effect on software effort estimation accuracy using moving windows. The existence of the Bellwether was empirically proven based on six postulations. We apply statistical stratification and Markov chain methodology to select the Bellwether moving window. The resulting Bellwether moving window is used to predict the software effort of a new project. Empirical results show that Bellwether effect exist in chronological datasets with a set of exemplary and recently completed projects representing the Bellwether moving window. Result from this study has shown that the use of Bellwether moving window with the Gaussian weighting function significantly improve the prediction accuracy.*

## 1. INTRODUCTION

Software effort estimation (SEE) is a core activity in software development life cycle [1] specifically for scheduling and monitoring a new software project development. SEE is defined as the process of predicting the most realistic effort needed to complete the development of a software project. The core objective of SEE is to obtain a robust and accurate predictive model which has proven elusive over the past years. The prediction accuracy of a model depends on the training and validation sets as well as the relative consistency of the nature of dataset considered. According to Mair et al. [2], there exist difficulty in obtaining relevant training data for building effort estimation models since the appropriate ones might be proprietary. This constraint makes researchers rely on using all available historical projects (referred to as the *growing portfolio*) as the training set which might be very old and will not necessarily yield a better prediction accuracy for new projects. Other researchers rely on using open source and cross company projects made publicly available which does not always guarantee a better prediction accuracy for new projects. Such projects are heterogeneous and inconsistent since they are extracted from different sources and may not have much bug-fixing support as compared to proprietary projects. This biasness in the selection criteria of training sets significantly affects the prediction accuracy resulting in unreliable estimation results [3]. Thus, there is the need to select relevant and not necessarily old completed projects as the training and validation sets for SEE.

According to Lokan and Mendes [4] [5], selection of recently completed projects to be used as the training set (referred to as *moving window*) from a chronological dataset can improve the prediction accuracy of the SEE model built for a *new* project. The moving window methodology is based on the assumption that recently completed projects are likely to share similar characteristics with *new* projects. Recent studies [6]–[7][8] have been conducted to support the theory by Lokan and Mendes that *the use of moving windows as the training set can improve the prediction accuracy of SEE models*. Recent studies [6] [7] have shown that *the use of weighting functions on moving windows can further improve the prediction accuracy of SEE models*. This theory was postulated by Amasaki and Lokan [6] [7] and was empirically confirmed to indeed improve the prediction accuracy when using large-sized moving windows. These previous studies found SEE models built using recently completed projects (*moving window*) superior to models that use all available historical projects (*growing portfolio*).

In the domain of software defect prediction, a recent study by Krishna et al. [9] has shown the existence of exemplary projects (referred to as *Bellwether*) to be used as the training set for making accurate predictions on new project cases. This concept of using *Bellwether* projects was also considered in previous studies [10] [11] whereby relevant selection of weekly sales data was used for making a new item's annual sale. The existence of Bellwether in a given dataset is referred to as *Bellwether effect* [9] and the search process involved in obtaining the Bellwether is referred to as the Bellwether method [9] or Bellwether analysis [10]. The Bellwether method uses the Transfer learning theory [12] to sample the Bellwether.

Kocaguneli et al. [13] empirically and successfully validated the theory that *Transfer learning can be used to obtain relevant cross company projects for better prediction accuracy*. The underlying principle behind the use of Transfer learning to select relevant data for improved prediction accuracy is thus analogous to the assumption we want to formulate for moving window effort estimation. We argue that *the selection of relevant and recently completed chronological projects can further improve the prediction accuracy irrespective of using cross company or single company project data for predictive modeling*. This can be made possible if projects are subjected to rigorous data preprocessing [14] and applying the concept of Bellwether [9] and Markov chains [15] for relevant selection of recently completed projects (referred to as the *Bellwether moving window*).

In Markov chain modeling [15], the outcome of an event (*new* project) in an experiment depends on the previous event (recently completed project with defined transition probabilities). The concept of *Bellwether* [9] [10] together with Markov chains [15] will assist in investigating the *Bellwether effect* in chronological datasets.

In this study, we make the following contribution: To the best of our knowledge, this is the first study to

- investigate the existence of *Bellwether* in software effort estimation datasets
- investigate the significance of *Bellwether effect* on software effort estimation accuracy
- introduce the *Bellwether moving window* with defined window size and age parameters for effective predictive modeling.

The *window size* simply refers to the number of most recently completed projects appropriate for building the prediction model. The *window age* refers to the elapsed time of projects within the window that has existed for not more than t calendar years or calendar months [8]. This study is unique since it improves on existing moving windows constraint that different sizing and aging parameters of the moving window affects the prediction accuracy [7] [8] [6].

The remaining sections of the paper are organized as follows. Section 2 presents the details of the *Bellwether* concept and the Markov chain Monte Carlo approach. Section 3 presents the postulations with their respective proofs. The proposed approach for selecting the *Bellwether moving window* is presented in Section 4. Section 5 details the methodological procedure employed. Section 6 presents the experimental results and discussion from the empirical analysis of the study. Section 7 gives a summary of related works with regards to the use of moving windows for SEE. Section 8 presents the threats to validity and Section 9 gives a summary of the study based on conclusions and future directions.

## 2. BACKGROUND

### A. The Concept of Bellwether

According to Krishna et al. [9], the concept of *Bellwether* which is a Transfer learning technique is defined in 2-folds namely the *Bellwether effect* and the *Bellwether method*:

1. *The Bellwether effect states that given a set of historical projects, there exist an exemplary project(s) that can form the Bellwether for predicting the target of new projects.*

2. *The Bellwether method uses a heuristic approach to search for that particular Bellwether from a set of historical projects. The subset of projects with the best prediction accuracy over the remaining projects is considered as the Bellwether.*

In the domain of software defect prediction, Krishna et al. [9] investigated the existence of *Bellwether* in nonchronological datasets. Thus, given a set of N nonchronological projects, each project data was used as a potential *Bellwether* to successfully make predictions on the remaining *N-1* projects. They reported a *Bellwether* if that project yielded accurate predictions on most of the *N-1* projects. Results from their study show that *Bellwethers* improve the prediction accuracy of estimation models. To the best of our knowledge, this is the first paper to investigate the existence of *Bellwether* projects in software effort estimation. We discuss into details the *Bellwether method* considered for investigating the *Bellwether effect* in chronological datasets in Section 4. In summary, we first sort the historical projects based on their completion dates and apply a statistical stratification technique to obtain only the recently completed projects (*moving window*). We then check for the existence of *Bellwether* by sequentially using each potential *Bellwether* subset to make predictions on the remaining projects. Thus, the *Bellwether* subset or window, $w_i$ is used to estimate each of the remaining windows, $w_j$ in the *N* projects, $\forall\ i \neq j$. The *Bellwether* subset with the best prediction accuracy is defined as *Bellwether moving window* and used to predict the effort of the new project.

### B. The Moving Window Concept

The moving window concept involves the use of recently completed projects with completion dates less than the start date of a new project. This concept has been a focus for most researchers [6] [8] [7] [16] over the past years. Weighting functions have been proven to further improve the prediction accuracy in a recent replication study by Amasaki and Lokan [6]. They introduced four weighting functions namely Triangular, Epanechnikov, Gaussian and Rectangular (or uniform) as shown in Fig. 1. It is worth noting that the Rectangular function is equivalent to unweighted moving window [6] whilst the other three functions are for the weighted moving windows. This paper makes use of the four weighting functions on the *Bellwether moving window* to further investigate the prediction accuracy in SEE modeling.

| Weighting function | Formula |
|---|---|
| Rectangular (Uniform) | $w(x) = 1$, $|x|<1$ |
| Triangular | $w(x) = 1 - |x|$, $|x|<1$ |
| Epanechnikov | $w(x) = 1 - x^2$, $|x|<1$ |
| Gaussian | $w(x) = \exp(-2.5(x)^2/2)$ |

Figure 1. Weighting functions

**C. Markov Chain Monte Carlo**

The Markov chain Monte Carlo (MCMC) is a probabilistic technique that seeks to solve the problem of sampling by exploring a given sample space. This is done through the construction of an ergodic Markov chain (EMC) whose limiting distribution is the target distribution [15]. MCMC utilizes Markov chains to effectively simulate a variable X whose future states are independent of past states given the present state. MCMC has been applied and proven successful in different software engineering domains such as software testing [17], image processing [18] as well as applied mathematics, statistical and biomedical engineering.

This paper provides a statistical investigation on the use of MCMC to obtain a potential subset of projects to be considered as the *Bellwether*. Let D denotes a set of projects from a given repository (sample space) with d degrees of dimension (features). Then, $X = (X1, ..., Xd)\epsilon D$ can represent a sample whose limiting distribution is known and can form the *Bellwether*. Each of the sample points $X1, ..., Xd$ of the sample, X is allowed to take a discrete or continuous time value which denotes the project ages already known from the given repository (D). The Markov chain is then executed a number of times until the chain converges to its limiting distribution to obtain the target sample subset [15]. The selected sample with its respective s states (or project ages) and constructed limiting distribution forms the *Bellwether* which can be used as the moving window.

**D. Central Limit Theorem and Law of Large Numbers**

The MCMC relies on the fact that, the limiting properties of the ergodic Markov chain possesses some similarities of independent and identically distributed (*iid*) sequences [15]. Hence, in order to prove the existence of a *Bellwether* sample from a state space, the central limit theorem and the law of large numbers should hold. The central limit theorem states that as the sample size turns to infinity, the sample mean will follow a normal distribution with mean, $\mu$ and variance, $\sigma^2$ [19]. The law of large numbers states that as the sample size increases, its mean, $\bar{x}$ will be approximately the same as the population mean, $\mu$ [20]. The larger the sample size ($n > 30$), the smaller the sample variance, $s^2$ and turns to follow a normal distribution [21]. Although Markov chains are not independent, the law of large numbers and central limit theorem disprove that fact, thereby for ergodic Markov chains, successive execution between visits to the same states are independent [15]. Thus, a moving window sample, $X_i$ can be used as a *Bellwether* with defined window size and window age for a successive prediction of *new* projects.

**E. Conditional Probability**

The conditional probability of an event X is the probability that X will occur given the knowledge that an event Y has already occurred. This conditional probability of X given Y is denoted as $P(X|Y)$ and defined as the ratio of $P(X \cap Y)$ to $P(Y)$. We define events X and Y as the sample subsets chronologically drawn from a sample space of projects and can form potential *Bellwethers* or training sets for predictive modeling. In the case whereby X and Y are independent with respect to prediction probabilities of the target or new projects, then $P(X \cap Y) = P(X)P(Y)$. This will result in $P(X|Y)=P(X)$ and conversely, $P(Y|X) = P(Y)$. Thus, if prediction is made from the probability of X given Y and assuming that the prediction probabilities of X and Y are independent, then $P(X|Y)$ is the same as $P(X)$. We define the *prediction probability* as the ratio of the predicted outcomes from a prediction model to the actual values of the data used for modeling.

**3. POSTULATIONS**

In this study, we prove the following postulations based on theoretical and empirical analysis that:

1. *If $\Omega$ is a sample space of a set of chronological projects, then the prediction probabilities of its partition sample sets $\{X_1, ..., X_q\}$ are independent.*

*Proof:* Suppose the sample space, $X$ is partitioned into q partition sample sets whereby each $X_i \epsilon \Omega$ is drawn independently, then the prediction probabilities of two events, $X_i$ and $X_j$ can be found as follows:

Assume that $X_i$ and $X_j$ are mutually exclusive (that is, the two events do not occur at the same time), then according to the conditional probability rule,

$$P(X_i|X_j) = \frac{P(X_i \cap X_j)}{P(X_j)} \quad (1)$$

$$P(X_i|X_j)P(X_j) = P(X_i \cap X_j)$$

$$P(X_j|X_i) = \frac{P(X_i \cap X_j)}{P(X_i)} \quad (2)$$

$$P(X_j|X_i)P(X_i) = P(X_i \cap X_j)$$

provided that $P(X_i) > 0$ and $P(X_j) > 0$.

To show that the events $X_i$ and $X_j$ are independent, then the prediction probability of $X_i$ does not affect the prediction probability of $X_j$ and vice versa. Thus, $P(X_i|X_j) = P(X_i)$ and $P(X_j|X_i) = P(X_j)$.

From the multiplication rule which states that, the probability that two events $X_i$ and $X_j$ occur is the product of the probability of $X_i$ and the probability of $X_j$.

That is,

$$P(X_i \cap X_j) = P(X_i)P(X_j) \quad (3)$$

Substituting (3) into (1) gives

$$P(X_i|X_j) = \frac{P(X_i)P(X_j)}{P(X_j)} = P(X_i) \quad (4)$$

Similarly, substituting (3) into (2) gives

$$P(X_j|X_i) = \frac{P(X_i)P(X_j)}{P(X_j)} = P(X_j) \quad (5)$$

From (4) and (5) it is confirmed that events $X_i$ and $X_j$ are independent. Similarly, from (1) and (2), it can be seen that

$$P(X_i|X_j)P(X_j) = P(X_i \cap X_j) \text{ and } P(X_j|X_i)P(X_i) = P(X_i \cap X_j)$$

Which implies

$$P(X_i \cap X_j) = P(X_i)P(X_j) \text{ since } P(X_i|X_j) = P(X_i).$$

Generally, $P(X_i \cap ... \cap X_q) = P(X_i) ... P(X_q)$ which implies that the prediction probabilities of the partition sample sets are independent.

We validated this postulation based on empirical analysis whereby each partition subset from the sample space resulted in different prediction probabilities of the target variable (software effort of the project). For example, in order to estimate the efforts, $Y$ of new projects (test set) using two samples (training sets), $X_1$ and $X_2$ from the ISBSG dataset, we realized that their respective prediction probabilities are as follows: $P(X_1) = 0.69$ and $P(X_2) = 0.43$. Thus, the training set, $X_1$ can predict about 70% of the correct efforts of the training set whiles $X_2$ can predict about 40% when such samples are used in prediction modeling.

2. *Given that the independent partition sample sets $\{X_1, ..., X_q\}$ have respective means$\{\overline{x}_1, ..., \overline{x}_q\}$, then there exist a sample mean, $\overline{X}$ which can be used as an unbiased estimate of the population mean, $\mu$.*

*Proof:* Let $X_1, ..., X_q$ be an independent and identically distributed (*iid*) sequence with finite means, $\overline{x}_1 < \infty, ..., \overline{x}_q < \infty$ respectively, then according to the strong law of large numbers [15] [20], as the sample size increases then there exist a sure probability of one [15] that

$$\lim_{q \to \infty} \frac{x_1 + x_2 + \cdots + x_q}{q} = E\left(\frac{\Sigma X_i}{q}\right) \approx E(\overline{X}) = \mu \quad (6)$$

Thus, according to Dobrow [15], the limit of the average of samples, $X_i$ as the number of samples, q approaches infinity is the expectation of the sample mean which can be used as the population mean. Similarly, the confidence interval of the sample. $X$ can be computed as shown in (7).

$$\overline{X} \pm Z_{\alpha/2}\left(\frac{S}{\sqrt{q}}\right) \quad (7)$$

where $\overline{X}$ is the sample space mean, S is the standard deviation of the sample space, $q$ is the number of independent partition subsets in the sample space and $Z_{\alpha/2}$ is the value from the standard normal distribution based on the desired confidence level ($\alpha$). Confidence interval computation assist in verifying the amount of uncertainty associated with the sample mean estimate to the population mean parameter. The lower and upper limits of the confidence interval forms the range for $\overline{X}$ and can be used to infer the range the population mean, $\mu$.

3. *If the sample space follows a normal distribution, then there exist a partition sample, $X_i$ that can be used as the Bellwether.*

Proof: Let the sample space in which the respective sample ($X_i$) to be drawn from, follow a normal distribution with mean, $\overline{X}$ and variance, $S^2$. That is, $X_i \sim N(\overline{X}, S^2)$. To achieve this assumption, we apply the *log transform*, the *z-score* normalization and other preprocessing techniques such as the Q-Q plot, kernel density plot [22] and the Cook's Distance [23] for treatment of outlier and influential data points in the sample space. We investigated if the sample (Xi) followed the normal distribution based on the skewness and kurtosis metrics. For normality, the skewness and kurtosis of Xi should be close to zero and three respectively [24]. Also for normality, as the sample size increases, its sample mean will follow the normal distribution (Central limit theorem [19]). Results from our empirical analysis show that the *log transform* converted the features or attributes of the sample space to follow the normal distribution with skewness approximately zero and kurtosis approximately three (see results in Section 6). We also realized that as the size of the sample space turns to be more than 100, the normality assumption is not only met but there exist a *Bellwether effect* in such a sample space.

For $\{X_1 ..., X_q\} \epsilon X$, then $\exists X_i$ such that: $P(X_i) \approx P(X)$. If $P(X_i) \approx P(X) = \frac{predicated\ outcome}{actual\ outcome}$ then $X_i$ follows a normal distribution with mean, $\overline{X}_i$ and variance, $S_i^2$. That is, $X_i \sim N(\overline{X}_i, S_i^2)$. The $X_i$ with the best prediction accuracy among the remaining $X_j \forall j \neq i$ forms the *Bellwether* sample yielding approximately similar prediction accuracy as compared to the overall sample space, $X$. We validated this postulation based on empirical analysis (see Section 6).

4. *If the partition samples $\{X_1, ..., X_q\}$ have unknown distributions, then there exist a particular sample, $X_i$ following the normal distribution that can be used as the Bellwether.*

Proof: Suppose that, the mean and variance of the partition sample, $X_i$ can be found and $X_i$ can make successful predictions of $X_j \forall j \neq i$ then Xi can be assumed to follow the Bernoulli distribution with a success prediction probability of $p$ and a failure prediction probability of $1 - p$. That is, $X_i \sim Ber(l, p)$ where the expected value of $X_i$ is $E(X_i) = p$, variance of $X_i$ is $Var(X_i) = p(1 - p)$ and the number of experimental outcome is 1 (success or failure).

If the prediction probabilities of $\{X_1 ..., X_q\}\epsilon X$ are independent of each other, then $X$ can be assumed to follow the Binomial distribution. That is, $X \sim Bin(n, p)$ where the expected value of $X$ is $E(X) = np$ variance of $X$ is $Var(X) = np(1 - p)$ and the number of experimental outcomes is $n$ (total number of prediction outcomes of all the q partition samples).

Now, following the Normal approximation to the Binomial distribution [25] [26], when $X$ follows the Binomial distribution with mean or expected value of $np$, variance of npq and when $n$ is large, then $X$ can approximately follow the normal distribution with mean, $\mu = np$ and variance, $\sigma^2 = np(1-p)$. That is, $X \sim N(\mu = np, \sigma^2 = np(1-p))$.

5. *If P denotes a regular transition probability matrix (TP M) of a Markov chain, then there exist an ergodic Markov chain (EMC) whose respective sample can be used as the Bellwether.*

*Proof:* We define a Markov chain as a set of random variables $\{X^{(1)}, ..., X^{(t)}\}$ or a collection of stochastic events $\{X(t)|t \geq 0\}$ whereby given the present event of a state at time t, the prediction of the future event at $t + 1$ is independent of the past events. Here, the sample subsets are composed of projects with ages $(t)$. The independent and identically distributed ($iid$) sample subsets $\{X^{(1)}, ..., X^{(t)}\}$ for all states at their respective times can be described as a Markov chain if

$$P(X_{t+1} = i_{t+1}|X_t = i_t, X_{t-1} = i_{t-1} \cdots X_1 = i_1, X_0 = i_0) = P(X_{t+1} = i_{t+1}|X_t = i_t) = p_{ij}^k \quad (8)$$

Assume there exist a variable 'P' that can be formulated as a matrix of transition probabilities of a Markov chain [15] from the sample space $\{X^{(1)}, ..., X^{(t)}\}$ where $i \epsilon T$ denotes the transition states (project ages), then the $ij^{th}$ element, $p^{(k)}ij/\epsilon P$ is the probability that the Markov chain starting from a particular state, $t_i$ will transition to $t_j$ after $k$ steps. If $p^{(k)}ij$ is homogeneous (or regular), then (9) holds and there exist a unique probability matrix, $\theta_u$ that can form the ergodic Markov chain (EMC) such that for any $\theta_o$ and for large values of u (10) can be defined as:

$$P(X^{(k)} = j|X^{(k-1)} = i) = p_{ij}^{(k)} \quad (9)$$

$$\lim_{u \to \infty} \theta_{u+1} = P^u \theta_1 \quad (10)$$

Thus, the ergodic Markov chain (EMC) can be obtained given a non-negative power of the transition probability matrix (TPM) by making all the entries of the probability matrix, $P'$ non-zero and irreducible. If the limiting state probability matrix (EMC) exist from the TPM constructed with the moving window sample, then that particular sample can be used as the *Bellwether moving window*. We consider such moving window sample as *Bellwether* since its limiting or stationary distribution has been reached and hence can form a potential training set for modeling given it has the best accuracy measures.

6. *Given a Bellwether whose ergodic Markov chain is known, then its size and age can be defined.*

*Proof:* Let a sample X with projects $\{p_1, ..., p_n\}$ be classified into $t_i$ states based on their respective ages. Assume the projects $\{p_1, ..., p_n\}$ from a particular sample is sorted in either non- increasing or non-decreasing order based on the project ages and the EMC of the sample is known. Then, the age of the *Bellwether* can be found as the difference between the maximum and minimum states (ages) of the sample whose EMC is known. Similarly, the size of the *Bellwether* can be found as the total number of projects of the sample.

## A. Assumptions

The following assumptions are made for the aforementioned postulations and the effective functioning of the proposed *Bellwether method* in section 4:

- Each project used in modeling has a start and completion date. Thus, we pruned off irrelevant projects prior to modeling.
- All selected projects from N have the same features.
- All selected projects share common development policies or characteristics.
- Each $pij$ element in the transition probability matrix lies within 0 and 1. Thus, $0 \leq pij \leq 1$.
- The sum of each $i^{th}$ row of the transition probability matrix should be 1.
- All entries of the ergodic Markov chain are non-zero.

## 4. BELLWETHER METHOD IN CHRONOLOGICAL DATASETS

In order to support the replication of this study, we describe our procedure in obtaining the *Bellwether* from chronological datasets in this section. Thus, we provide a detailed step-by-step procedure to first find the existence of *Bellwether* in the utilized datasets and how they can be used as the training sets for estimating the software effort of new projects. Given a new project with an unknown software effort, select relevant projects to be used as the *Bellwether* from a historical dataset (D) using the following three main operators (*SORT+CLUSTER, GENERATE TPM and APPLY*):

### SORT + CLUSTER
*Given a set of N historical and completed projects from D, sort based on the project completion dates and stratify the data into q clusters or windows.*

1. For all *N* projects from *D* sort in an increasing order using their respective project completion dates.

2. Subject sorted data to $X - means^1$ clustering algorithm [27] to obtain *q* clusters. Perform data stratification based on the *q* obtained. That is, stratify *N* into q clusters whereby each cluster (or window) has a set of chronologically arranged projects.

3. For each cluster or window, compute the weighted moving window by applying the respective weighting functions (see Fig. 1 in Section 2) on all the *q* windows. We define the resulting q weighted windows in an increasing order as $w_1, w_2 ... w_q$.

4. Use $w_q$ as a baseline window. It should be noted that, $w_q$ contains recently completed projects in a chronological order and can form the *baseline weighted moving window*.

### GENERATE TPM
*Generate the transition probability matrix (TP M) for the resulting weighted moving window and find the respective ergodic Markov chain (EMC).*

---
[1] X-means automatically estimate the number of optimal clusters with respect to the Bayesian Information Criterion [27]

5. For the resulting weighted moving window ($w_q$), generate the TPM. Here, use the project ages of the moving window as the transition states in order to generate the TPM.

6. Compute the EMC for the generated TPM in order to validate the limiting (or stationary) distribution of $W_q$. Thus, using the TPM with $t_i$ states, successively perform the squaring of TPM until TPM becomes regular or reaches its stationary distribution (EMC). That is, when individual probability elements, $p_{ij}$ of TPM cannot be reduced further, we say that TPM is regular or ergodic [15].

7. Report $w_q^*$ as a *Bellwether moving window* if the weighted moving window, $w_q$ is stationary and provides a better prediction accuracy of majority of the remaining weighted moving windows, $w_1, w_2 \dots w_{q-1}$. Else go to step 4 to update wq with additional project(s) from $w_{q-1}$ and repeat the GENERATE TPM operator. Conversely, update $w_q$ by removing project(s) from wq and adding to $w_{q-1}$ and repeating the GENERATE TPM operator.

8. Evaluate the accuracy of $w_q^*$ using effective performance measures. It should be noted that, we used Mean Absolute Error (MAE), Mean Balanced Relative Error (MBRE) and Mean Inverted Balanced Relative Error (MIBRE) as suggested by Foss et al. [28].

*APPLY*

Apply the Bellwether moving window, $w_q^*$ to the new project data (set as a hold-out) whose software effort is to be estimated.

9. Once $w_q^*$ is obtained as a *Bellwether moving window* in step 7 prior to its application to the *new* project data (set as a hold-out), its size and age can be defined from its dimensions. The size of the moving window is computed as the total number of projects within the *Bellwether moving window* whiles the age of the moving window is computed as the difference between the maximum and minimum project ages.

10. Predict the software effort of the *new* project using the resulting *Bellwether moving window*.

## 5. METHODOLOGY

**A. Dataset Description**
In this study, we use the International Software Benchmarking Standards Group (ISBSG) dataset release 10 which is sourced from the ISBSG dataset repository[2] and the Kitchenham dataset also sourced from the *tera-Promise* repository[3].

To facilitate an effective investigation of the existence of *Bellwether* in SEE, we used the same ISBSG dataset and Kitchenham dataset that have previously been considered for moving window studies [7] [4], [8] [29].

The ISBSG dataset is a chronological cross company dataset with a total of 4106 projects. It contains variants of projects from different organizations. All projects are within a range of May, 1988 to November, 2007 with a span of 20 years approximately. We found that new development projects tend to require higher efforts than enhancement projects. There is also no unique pattern with respect to specific programming languages used for development. Further details of the ISBSG datasets are provided in previous studies [7] [8] [30].

The Kitchemham dataset is a chronological dataset which was first applied in moving windows by Kitchenham et al. [29]. The dataset is made up of 145 projects from a single company and falls within a range of 1994 to 1999. Further details about this dataset are provided in a study by Kitchenham et al. [29].

Although the aforementioned datasets seem old, the focus is not how old the datasets are but to prove that new projects can be predicted from previously completed projects based on the *Bellwether effect* and Markov chains.

**B. Data Preprocessing**
We preprocess the ISBSG dataset following a similar approach by previous studies [7] [8]:

- Eliminate projects if their respective ages (elapsed durations) are unknown.
- Eliminate projects with low data quality rating (that is, projects not rated as A or B by ISBSG).
- Eliminate projects with outdated function points (that is, projects whose IFPUG versions are below 4.0)
- Eliminate projects with unknown development team efforts.
- Eliminate projects whose unadjusted function point sizes are unknown.
- Eliminate projects which are considered as web projects.
- Eliminate projects with missing data values.
- Cook's distance [31] is employed to identify and remove influential projects during model construction following a similar approach by Lokan and Mendes [8]. Thus, projects that did not contribute significantly to fitting the prediction model are removed.
- Apply data transformation technique (*log transform*) to the ratio-scaled features (eg. size (UFP), effort, etc).

The preprocessing resulted in a total of 1097 projects (26.7%) selected from the population set (4106 projects).

The Kitchenham dataset was preprocessed following a similar approach by Kitchenham et al. [29] whereby projects with the actual efforts and function point count were retained. All projects without estimated completion dates were eliminated. As a result of the heterogeneous and inconsistent nature of the dataset, influential data points which did not contribute significantly to model building were eliminated. Out of the total *of* 145 projects from the

---

[2] http://www.isbsg.org

[3] http://openscience.us/repo/

Table 1. Descriptive Statistics of Preprocessed Ratio-scaled Features

| Feature | Mean | Std Dev | Min | Max |
|---|---|---|---|---|
| ISBSG Dataset | | | | |
| PDR | 0.575 | 0.047 | 0.461 | 0.691 |
| Size (UFP) | 0.545 | 0.249 | 0.106 | 0.812 |
| Elapse_time | 0.544 | 0.023 | 0.420 | 0.624 |
| Effort | 0.732 | 0.052 | 0.554 | 0.927 |
| Kitchenham Dataset | | | | |
| AFP | 0.663 | 0.036 | 0.573 | 0.816 |
| Elapse_time | 0.647 | 0.020 | 0.598 | 0.704 |
| Effort | 0.724 | 0.037 | 0.545 | 0.895 |

*PDR denotes the normalized Productivity Delivery Rate in hours per function point*
*UFP denotes the normalized Unadjusted Function Points*
*AFP denotes the normalized Adjusted Function Points*
*Std Dev denotes Standard Deviation*

Kitchenham dataset, 142 projects (97.9%) were obtained after preprocessing.

The selected features from the ISBSG dataset are the size of the projects measured in Unadjusted Function Points (UFP), primary language type (3GL, 4GL), development type (new development, re-development and enhancement), platform (PC, mainframe, midrange and multi-platform), industry sector (manufacturing, banking, insurance and other) and the effort measured in hours. On the other hand, the selected features from the Kitchenham dataset are Adjusted Function Points (AFP) and development effort measured in hours. With the exception of the effort regarded as the dependent or target variable, the remaining features are regarded as the independent variables following a similar procedure in previous studies [7], [8] [29]. We provide the following summary statistics of the ratio-scaled features (Table 1) using the *log transform* function on each dataset. The project size, effort, elapsed time measured in calendar years (ISBSG data) or days (Kitchenham data) and the Project Delivery Rate (PDR) measured as the ratio of the effort to the project size provided a foundation in proving the aforementioned postulations (Section 3). PDR was examined in a previous study [4] and it was found that it changes with time and low PDR suggests high productivity. This finding supports the use of moving windows in predictive modeling [7].

## C. Experimental Setup

We first investigate the existence of *Bellwether* in chronological datasets based on the aforementioned postulations in Section 3. This is done by conducting empirical experiments using the step-by-step procedure detailed in Section 4. Since this study seeks to improve the *moving window* modeling approach, we make use of the four weighting functions in our modeling setup as elaborated in step 3 of Section 4.

Comparison is made across the use of weighted moving windows (Triangular, Epanechnikov and Gaussian) and unweighted moving windows (Rectangular). It should be noted that, these weighting functions have been used in previous studies [7] [6] [32]. For the moving window, we evaluate its accuracy performance by using the selected window sample to perform successive predictions on the remaining unselected windows (see detailed explanation in Section 4). Thus, while the selected moving window is used as the training set, the unselected windows are used as the validation sets. Even though this approach shares similar characteristics with the *k-fold* cross validation [1], the approach differs since each kth partition[4] or sample might not necessarily have the same sample size as the other partitions. Thus, after obtaining the *Bellwether moving window* (partition sample with the best prediction accuracy on the remaining partitions), it is then used for estimating the new project (set as hold-out in each dataset). On the other hand, we used the *growing portfolio* (that is all preprocessed projects) as the training set as done in previous studies [7] [8] [6] and used the leave-one-out validation approach [1] to set up the prediction model. Thus, at each run N−1 projects are used for training and the remaining project for validation. Lastly, we compare the *Bellwether moving window* approach with the *growing portfolio* approach which has been a benchmark practice in previous studies [7] [6]. These two windows are used as the training sets for the prediction or learning models.

### 1) Prediction Models

Three prediction models (or learners) have been considered in this study, namely multiple linear regression (MLR) model, Automatically Transformed Linear Model (ATLM) and a Deep learning model.

Previous studies using moving windows in modeling [7] [6] have shown that the multiple linear (or ordinary least squares) regression yielded a better prediction accuracy than complex models (eg. Analogy Based Estimation) when weighted moving windows and growing portfolio were considered as the training sets for the prediction model. Thus, since this study aims at investigating moving windows based on the *Bellwether effect*, we consider modeling with multiple linear regression (with the same dependent and independent variables [7] [29]) relevant so replication of previous studies [7] [29] results can be achieved.

The ATLM was proposed by Whigham et al. [1] as a baseline model in effort estimation. Whigham et al. argue that since there is no accepted standard for setting up a prediction model in SEE domain, their proposed baseline approach (ATLM) supports the *k-fold* cross validation, leave-one-out validation and a relevant sample size with repeated training and validation sets. They [1] compared ATLM to other complex models such as ensemble methods [3] and a hybrid approach (incorporating particle swarm optimization, analogy-based estimation and clustering [2]). ATLM yielded a better prediction performance as compared to the other prediction models and was considered as a baseline model for SEE [1].

We consider a complex prediction model namely a Deep learning model which has yielded better prediction accuracy in previous studies [33] [34], [35] [36]. Specifically, we construct a Deep Neural Network which makes use of multiple hidden layers and an output layer with their respective neurons to automatically learn from a set of projects and gives the resulting prediction for the target (in our case, the software effort of new projects). The Levenberg-Marquardt backpropagation optimization [37] training function is employed to update the weights of the neurons in the hidden and output layers respectively. The hyperbolic tangent activation function is used in each of the neurons for giving the respective outputs.

---
[4] The k was obtained based on X-means clustering algorithm

**2) Performance Measures**

We employ three performance measures which have been proven reliable by Foss et al. [28]. These are Mean Absolute Error (MAE), Mean Balanced Relative Error (MBRE) and Mean Inverted Balanced Relative Error (MIBRE). MAE have been considered in previous studies [1] [6] [38] to evaluate the accuracy of prediction models. Similarly, MBRE and MIBRE have been considered as effective evaluation measures in a study by Kocaguneli et al. [13].

MAE is a risk function that measures the average absolute deviation of the estimated effort (EE) values from the actual or true effort (EA) values and it is defined in (11). MBRE defined in (12) is the average of the ratio of the absolute deviations of the estimated effort (EE) from the true effort (EA) to the minimum values of the estimated effort and the true effort. Conversely, MIBRE defined in (13) is the average of the ratio of the absolute deviations of the EE from the EA to the maximum values of the EE and EA. Minimum values from the performance measures are considered superior in terms of model accuracy.

$$MAE = \frac{1}{n}\sum_{i=1}^{n}|E_{Ai} - E_{Ei}| \quad (11)$$

$$MBRE = \frac{1}{n}\sum_{i=1}^{n}\left(\frac{|E_{Ai} - E_{Ei}|}{\min(E_{Ai}, E_{Ei})}\right) \quad (12)$$

$$MIBRE = \frac{1}{n}\sum_{i=1}^{n}\left(\frac{|E_{Ai} - E_{Ei}|}{\max(E_{Ai}, E_{Ei})}\right) \quad (13)$$

**3) Statistical Tests**

A robust statistical test, the Welch t−test statistic recommended by Kitchenham [22] is used to determine the statistical pairwise differences among the weighted (Triangular, Epanechnikov and Gaussian) and unweighted (rectangular) windows. We considered the Kruskal-Wallis H−test statistic [39] to find the existence of statistical difference among the four weighting functions applied on the moving window. Kruskal-Wallis H−test which is a non-parametric test statistic was used by Krishna et al. [9] in making multiple comparison tests across two or more groups when examining the *Bellwether effect*. Statistical significant differences are considered at α=0.05 asymptotic significance level. The Glass' Δ effect size [40] is used to find the practical significance of the *Bellwether moving window* in chronological datasets. Effect size computation was considered in a previous study by Amasaki and Lokan [6] in relation to the preference of a particular moving window. A minimum effect size threshold value of 0.5 [40] [6] is used for selecting the effective window. We considered these performance measures due to their robustness to outliers.

## 6. RESULTS AND DISCUSSION

In this section, we present the empirical evidence of the existence of *Bellwether moving windows* in chronological datasets using the aforementioned postulations and the step-by-step procedure in Section 3 and 4 respectively.

### A. Bellwether Effect in Chronological Datasets

We first subjected each preprocessed and sorted dataset to the *X-means* clustering algorithm. Out of the total of 1097 (26.7%) sorted and preprocessed ISBSG dataset, 5 partition samples each with an approximate size of 219 projects (20.0%) were obtained. Similarly, out of the total of 142 (97.9%) sorted and preprocessed Kitchenham dataset, 3 partition samples each with an approximate size of 47 projects (33.1%) were obtained.

Table 2: Descriptive statistics of z-score normalization and log transform of partition samples (ISBSG Dataset)

| Sample | Features | | | | | | | |
|---|---|---|---|---|---|---|---|---|
| | PDR | | UFP | | Elapse time | | Effort | |
| | Skewness | Kurtosis | Skewness | Kurtosis | Skewness | Kurtosis | Skewness | Kurtosis |
| *log transform* | | | | | | | | |
| $X_1$ | -0.614 | 2.211 | -0.153 | 3.647 | -0.619 | 3.474 | 0.010 | 3.023 |
| $X_2$ | -0.261 | 3.344 | -0.226 | 1.874 | 0.011 | 2.992 | -0.223 | 3.115 |
| $X_3$ | -0.022 | 2.700 | -0.012 | 1.758 | 0.027 | 2.906 | -0.038 | 2.588 |
| $X_4$ | 0.091 | 3.118 | 0.147 | 1.930 | -0.104 | 2.984 | 0.123 | 3.132 |
| $X_5$ | 0.569 | 3.276 | 0.456 | 3.361 | 0.039 | 3.252 | -0.080 | 2.951 |
| *z-score normalization* | | | | | | | | |
| $X_1$ | 0.849 | 2.275 | -1.097 | 4.776 | -2.853 | 5.625 | 3.832 | 8.472 |
| $X_2$ | -0.885 | 4.878 | -0.097 | 1.777 | 1.669 | 5.234 | -0.283 | 7.563 |
| $X_3$ | 1.930 | 6.029 | -0.243 | 1.806 | -1.969 | 3.877 | 5.303 | 5.369 |
| $X_4$ | 5.482 | 7.818 | 0.053 | 1.806 | -0.783 | 5.645 | 7.575 | 6.117 |
| $X_5$ | 4.207 | 9.135 | 4.114 | 8.950 | 3.207 | 5.135 | 3.538 | 8.962 |

Table 3: Descriptive statistics of z-score normalization and log transform of partition samples (Kitchenham Dataset)

| Sample | Features | | | | | |
|---|---|---|---|---|---|---|
| | AFP | | Elapse time | | Effort | |
| | Skewness | Kurtosis | Skewness | Kurtosis | Skewness | Kurtosis |
| *log transform* | | | | | | |
| $X_1$ | -0.669 | 3.033 | 0.452 | 2.371 | -0.021 | 2.954 |
| $X_2$ | 0.497 | 4.532 | -0.136 | 3.475 | 0.158 | 3.088 |
| $X_3$ | 0.126 | 2.337 | 0.181 | 2.362 | 0.369 | 2.858 |
| *z-score normalization* | | | | | | |
| $X_1$ | 1.243 | 2.647 | 1.880 | 6.695 | 3.122 | 5.106 |
| $X_2$ | 3.159 | 9.298 | 2.136 | 7.299 | 2.208 | 9.708 |
| $X_3$ | 2.219 | 7.360 | 1.053 | 4.071 | 3.611 | 8.898 |

Results from Tables 2 and 3 present the partition samples, skewness and kurtosis values for the application of two data transformation techniques (*log transform* and *z-score normalization*) with regards to the ISBSG and Kitchenham datasets respectively. Results show that, when the various partition samples were transformed using the *log transform* and *z-score normalization*, it was observed that the *log transform* yielded skewness values of zero approximately whiles the kurtosis values were in an approximate range of 2 to 4. It should be noted that for a sample to follow a normal distribution, the skewness and kurtosis values should be around 0 and 3 respectively [24]. The application of the log transform forms the baseline for the existence of Bellwether effect, since it transformed the selected data with unknown distribution to follow the normal distribution (Postulations 3 and 4). Even though the z-score data transformation yielded approximate skewness and kurtosis values of 0 and 3 respectively, we realized that it was not across all partition samples (Tables 2 and 3). The empirical results do not only confirm the use of log transform in previous moving window studies [2] [7], [10] [12] [35] but also validate the baseline for finding the Bellwether moving window from partition samples which are normally distributed (Postulation 3). Each of the partition sample set, Xi is chronologically sorted and hence Xq from each dataset contains the recently completed projects (moving window). We therefore use the Xq in each dataset as the initial baseline window to search for the Bellwether moving windows (step 4 of Section 4). Here, q denotes the last partition in a chronological manner of sorting. That is, X5 for the ISBSG dataset (Table 2) and X3 for the Kitchenham dataset (Table 3). For the selected baseline windows from the two datasets, the respective transition probability matrices were generated and the existence of their respective ergodic Markov chains confirmed the stationarity of the baseline windows. Following the step-by-step procedure in Section 4, the baseline window (w∗q) was updated with projects from wq−1. Since the concept of Bellwether [9] involves systematic estimation of the remaining partition samples, we estimated and evaluated the prediction results of the remaining partition samples as illustrated in Section 4. After empirical analysis and evaluation, we realized that the resulting partition sample to be considered as the Bellwether moving window for the ISBSG dataset had the best window size of 257 projects (23.4%) and window age of 2.5 years approximately. Similarly, the Bellwether moving window for the Kitchenham dataset had the best window size of 87 projects (61.27%) and window age of 2 years approximately. This result slightly confirms previous studies [7] [8] [5] whereby they proposed the use of weighted moving window of relatively large size (more than 75 projects) and average window age of about 3 years for effective prediction performance.

We record the MAE, MBRE and MIBRE evaluation measures for all the 3 prediction models (leaners) in Tables 4, 5 and 6 respectively. Results from the MAE evaluation in Table 4 show that on average the application of the Gaussian weighting function (column (d)) yielded a relative effective prediction accuracy (minimum MAE highlighted in grey colour). This was observed for both the ISBSG and Kitchenham datasets as compared to the application of the three functions. Similar results were observed from the MBRE and MIBRE evaluation measures in Tables 5 and 6 respectively. Thus, even though Bellwether effect exist in the datasets, we realized a significant prediction improvement when the Gaussian weighting function was applied on the Bellwether moving window.

Table 4. MAE Evaluation of Bellwether Moving Window

| Learner | (a) | (b) | (c) | (d) |
|---|---|---|---|---|
| ISBSG Dataset | | | | |
| MLR | 6.0727 | 7.2388 | 4.4010 | 1.7645 |
| ATLM | 2.5912 | 3.7271 | 3.5912 | 1.6800 |
| DNN | 1.6974 | 1.9760 | 1.2316 | 0.5860 |
| Kitchenham Dataset | | | | |
| MLR | 2.6310 | 4.4698 | 4.7827 | 3.2336 |
| ATLM | 4.5652 | 2.1960 | 3.7827 | 1.1586 |
| DNN | 1.1408 | 1.1572 | 2.0986 | 1.0450 |

(a) Rectangular, (b) Triangular, (c) Epanechnikov, (d) Gaussian
MRL     Multiple Linear Regression
ATLM     Automatically Transformed Linear Model
DNN     Deep Neural Networks
*The grey shaded cells indicate the best (minimum) values across each learner*

Table 5. MBRE Evaluation of Bellwether Moving Window

| Learner | (a) | (b) | (c) | (d) |
|---|---|---|---|---|
| ISBSG Dataset | | | | |
| MLR | 4.9720 | 4.7114 | 3.3154 | 1.3020 |
| ATLM | 2.4378 | 3.4896 | 2.7832 | 1.5145 |
| DNN | 1.4962 | 2.0475 | 0.6816 | 0.7651 |
| Kitchenham Dataset | | | | |
| MLR | 4.1804 | 8.1908 | 3.8262 | 1.7791 |
| ATLM | 3.9130 | 5.0036 | 3.9130 | 0.8262 |
| DNN | 0.7628 | 2.4185 | 1.0595 | 0.5950 |

(a) Rectangular, (b) Triangular, (c) Epanechnikov, (d) Gaussian

Table 6. MIBRE Evaluation of Bellwether Moving Window

| Learner | (a) | (b) | (c) | (d) |
|---|---|---|---|---|
| ISBSG Dataset | | | | |
| MLR | 7.9916 | 3.2693 | 3.2471 | 1.3477 |
| ATLM | 4.4018 | 3.7669 | 3.1064 | 1.1586 |
| DNN | 5.1826 | 3.0877 | 1.6249 | 0.4725 |
| Kitchenham Dataset | | | | |
| MLR | 6.0055 | 3.2780 | 7.1057 | 3.0942 |
| ATLM | 3.3725 | 4.2464 | 3.2074 | 2.2241 |
| DNN | 2.0475 | 1.4952 | 2.0638 | 1.1024 |

(a) Rectangular, (b) Triangular, (c) Epanechnikov, (d) Gaussian

Table 7. MIBRE Evaluation of Bellwether Moving Window

| Performance Measures | Glass' Δ Effect size across Learners | | |
|---|---|---|---|
| | MLR | ATLM | DNN |
| ISBSG Dataset | | | |
| MAE vs. MBRE | 0.2647 | 0.1813 | 0.6470** |
| MAE vs. MIBRE | 0.1756 | 0.1518 | 0.7570** |
| MBRE vs. MIBRE | 0.9110** | 0.2650 | 0.9215** |
| Kitchenham Dataset | | | |
| MAE vs. MBRE | 0.0909 | 0.0395 | 0.9618** |
| MAE vs. MIBRE | 0.6683** | 0.2534 | 0.9927** |
| MBRE vs. MIBRE | 0.5774** | 0.2116 | 1.1029** |

*Practical Significance:* **Δ>0.5

With regards to the use of the 3 learners (MLR, ATLM and DNN), we realized on average that the DNN yielded the best prediction accuracy along each function for each evaluation measure (Tables 4 to 6). With regards to the Glass' Δ effect size [40] computation across the 3 learners (Table 7), we realized that the DNN model yielded practical significant effects as compared to the ATLM and MLR. Thus, at a threshold of 0.5 [6] [40], the effective size pairwise differences of the evaluation measures for DNN were significant.

> *Results show that, the Bellwether Effect exist in chronological datasets and its respective Bellwether moving window yields superior prediction accuracy when the Bellwether projects are not older than 2.5 years of existence.*

> *Results show that, the Gaussian weighting function applied on the Bellwether moving window is statistically significant with respective to relative effective prediction performance as compared to the other weighting functions.*

### B. Statistical Significance of Bellwether Effect with Respect to Weighting Functions

We performed statistical tests to investigate the significant differences across the four weighting functions, namely Rectangular, Triangular, Epanechnikov and Gaussian functions. This was done by computing the Kruskal-Wallis H− test [39] for each prediction modeling approach (Table 8). Results from the 3 prediction models (learners) show that, even though there *Bellwether* exist, the application of the weighting functions on the *Bellwether moving windows* resulted in significant differences at 5% significance level. This affected the prediction accuracy performances.

As a result of the existence of statistical significant differences across the application of the four weighting functions on the *Bellwether moving windows*, we further investigated the pairwise comparison of the weighting functions. This was done using the Welch t-test [22] and the Glass' Δ effect size [40] to confirm the statistical significance of the pairwise differences. Results from Table 9 show that there exist statistical significant differences in the application of the Gaussian weighting function in all cases. This confirms the results obtained in Tables 4, 5 and 6 that the application of the Gaussian weighting function on the *Bellwether moving window* yields significant prediction accuracy as compared to the other weighting functions. Similar result was found in a previous study [7] whereby the application of the Gaussian function on the moving window yielded significant effect on accuracy. For statistical and practical significance, the p− *value* and the effect size value should be less than 0.05 and greater than 0.5 respectively (Table 9).

Table 8. MAE Evaluation of Bellwether Moving Window

| Learner | Chi-square | p-value |
|---|---|---|
| **ISBSG Dataset** | | |
| MLR | 6.53 | 0.0311* |
| ATLM | 9.88 | 0.0435* |
| DNN | 3.03 | 0.0314* |
| **Kitchenham Dataset** | | |
| MLR | 6.42 | 0.0146* |
| ATLM | 4.35 | 0.0256* |
| DNN | 8.04 | 0.0414* |

Statistical Significance: *p<0.05

Table 9. MIBRE Evaluation of Bellwether Moving Window

| Weighting Function | Welch t-test | | Glass' Δ Effect size |
|---|---|---|---|
| | t-value | p-value | |
| **ISBSG Dataset** | | | |
| Gaussian vs. Rectangular | -3.8929 | 0.0039* | 1.3271** |
| Gaussian vs. Triangular | -4.6063 | 0.0011* | 1.6025** |
| Gaussian vs. Epanechnikov | -3.4130 | 0.0063* | 1.2194** |
| Rectangular vs. Triangular | 0.4350 | 0.6700 | 0.2488 |
| Rectangular vs. Epanechnikov | 1.7052 | 0.1129 | 0.1710 |
| Triangular vs. Epanechnikov | 1.5609 | 0.1393 | 0.4870 |
| **Kitchenham Dataset** | | | |
| Gaussian vs. Rectangular | -2.3892 | 0.0328* | 0.9247** |
| Gaussian vs. Triangular | -2.4399 | 0.0326* | 0.8929** |
| Gaussian vs. Epanechnikov | -3.0018 | 0.0105* | 1.1483** |
| Rectangular vs. Triangular | -0.4349 | 0.6698 | 0.1828 |
| Rectangular vs. Epanechnikov | -0.5773 | 0.5718 | 0.2660 |
| Triangular vs. Epanechnikov | -0.0732 | 0.9426 | 0.0389 |

Statistical Significance: *p<0.05. Practical Significance: **Δ>0.5

### C. Prediction by the Bellwether Moving Window and the Growing Portfolio

We compared the *Bellwether moving window* with the *Growing portfolio* (all preprocessed historical projects as training set) by setting a completed project as a hold-out from each dataset which was not used in the training and validation sets. Thus, we used both the *Bellwether moving window* and the *Growing portfolio* to estimate the software effort of the new projects from the 3 learners as shown in Fig. 2 and Fig. 3 respectively. It should be noted that these new projects already have known software effort (actual effort) from both datasets respectively. For example, we predicted the effort of a project developed in November, 2007 with previously developed projects from the ISBSG dataset (Fig. 2). Similar approach is followed for the Kitchenham dataset (Fig. 3). These known or *actual efforts* enabled us in making comparison with the *predicted efforts* (Fig. 2 and 3).

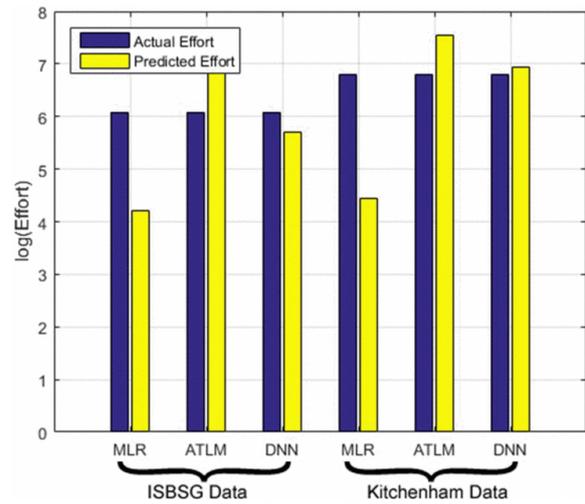

Figure 2. New project's software effort estimation using the Bellwether moving window

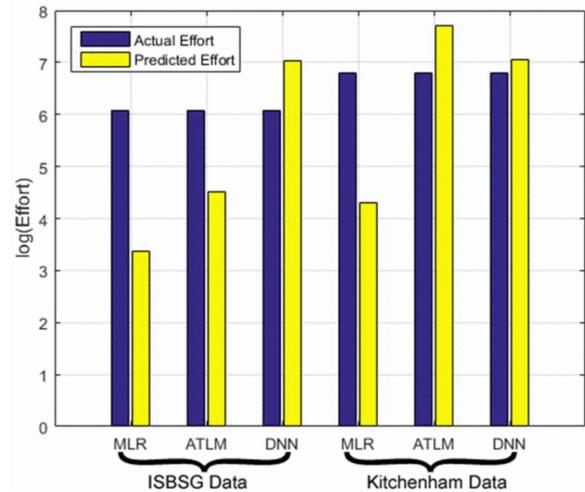

Figure 3. New project's software effort estimation using the Growing portfolio

The Deep learning (DNN) model trained with the *Bellwether moving window* (Fig. 2) resulted in relatively superior accuracy with respect to predicting the software effort of the new project than the other models in both datasets. In the case of the *Growing portfolio* (Fig. 3), we realized that the DNN also resulted in superior prediction accuracy similar to but not better than the *Bellwether moving window* in Fig. 2.

> *Results show that, the Bellwether moving window yields relative effective prediction when trained with a Deep learning model (DNN).*

## 7. RELATED WORK

To the best of our knowledge, Kitchenham et al. [29] were the first to consider the selection of recently completed projects (*moving windows*) from chronologically completed projects for SEE. They made use of the Kitchenham dataset comprising of chronological projects obtained from a single company. They divided the *growing portfolio* into four partitions based on the development start date and each partition was used as the training set for modeling. Results from their regression analysis show that the size to effort estimation changed across the four strata. They concluded that, older projects should be removed from the dataset and estimation done with the remaining subset together with addition of recently completed projects. Lastly, they recommended the estimation of new projects with respect to the use of about 30 recently completed projects.

A recent study by Amasaki and Lokan [6] investigated the effect *of* weighted moving windows on SEE accuracy using the Finnish dataset. In their study, they applied four weighting functions namely Triangular, Epanechnikov, Gaussian and Rectangular shown in Fig. 1 (Section 2). They realized that the application of different weighting functions have different effects on the SEE accuracy. They concluded that, weighted moving windows yielded statistically significant effect with comparably larger window size as compared to unweighted moving windows. Their findings confirmed results from their previous study [7] where similar results were found using the ISBSG dataset. This present paper applied the weighting functions to the *Bellwether moving windows*. It is worth noting that the Rectangular function [6] is equivalent to unweighted moving window whilst the other three functions are for the weighted moving windows.

Amasaki and Lokan [6] also compared weighted moving windows to the *growing portfolio* (using all historical projects as training set), they realized that the *growing portfolio* yielded a better significant improvement on the SEE accuracy. This is confirmed in a previous study by Lokan and Mendes [8] using the Finnish dataset. Amasaki and Lokan [6] considered window sizes ranging between 20 and 120 projects (ie. 20, 30, 40, 120) in building their estimation model based on weighted linear regression. Results show that models built with windows of sizes ranging from 20 to 30 projects yielded minimum Mean Absolute Error (MAE) as compared to the use of *growing portfolio* in SEE modeling. The prediction accuracy diminishes when using Triangular, Epanechnikov and Gaussian weighting functions for window sizes below 80 projects whiles the unweighted moving windows (rectangular) resulted in better prediction accuracy. Lastly, with window sizes more than 80 projects, the prediction accuracy is highly improved when using the three weighted moving windows (Triangular, Epanechnikov and Gaussian) as compared to using unweighted moving windows (rectangular).

A similar study by Amasaki and Lokan [7] examined the effects of weighted and unweighted moving windows on the prediction accuracy. They considered projects from the ISBSG repository (release 10) and used weighted linear regression in building the SEE model. Results from their study show that: 1) unweighted moving windows are more effective in SEE with larger window size (at least 40 projects) as compared to unweighted growing portfolio; 2) weighted growing portfolio of large size yielded better estimation accuracy as compared to unweighted growing portfolio with respect to MAE; 3) unweighted moving window is more effective in prediction accuracy than weighted growing portfolio; 4) for large window size of at least 100 projects, weighted moving window yielded better prediction accuracy than unweighted growing portfolio for both MAE and Mean Magnitude of Relative Error (MMRE). Here, the Gaussian weighting function was considered the best; 5) weighted moving windows of large sizes are better than weighted growing portfolio with respect to MAE and MMRE. The Gaussian function was considered the best weighting function in this case; and 6) Weighted moving windows with large sizes (at least 80 projects) are more effective in prediction accuracy than unweighted moving windows for MAE and MMRE. Here, no preference to the best weighting function was made in their study. They [7] concluded from their study that using windows of large size (more than 75 projects) improve the prediction accuracy.

Lokan and Mendes [8] investigated the effect of using moving windows based on different durations and how it affects the prediction accuracy. The ISBSG dataset (release 10) and the Finnish dataset were considered in their study. They realized that, the prediction accuracy can be improved when using moving windows that are based on duration. Here, stepwise multivariate regression was used in building the SEE model and windows of different durations ranging from 1 to 7 years were considered in modeling. Results show that SEE models built with moving windows of about 3 years with at least 75 projects (training set) yielded better significance in accuracy. Similar results were achieved in their previous study [5] whereby they found that moving windows of about 3 to 4 years yielded better prediction accuracy with the best window size *of* 81 to 89 projects from the ISBSG dataset.

The aforementioned studies discussed show that there is an implicit assumption that moving windows are useful in predictive modeling. Even though empirical analysis has shown that the moving windows approach respond differently across datasets, more studies need to be conducted to further improve the prediction accuracy. In this present paper, we have introduced the *Bellwether moving window* (with defined window size and age) which can be considered as the training set for predictive modeling of new projects.

## 8. THREATS TO VALIDITY

We used a single release of the ISBSG dataset (release 10) as considered in previous studies [7] [4] [8] and the Kitchenham dataset [29]. These datasets are convenient target population sets but cannot be a general representative of all chronological datasets. Therefore, results might not be confidently generalized beyond these datasets.

The models employed in this study were automated and hence can result in automation biasness. For example, the selection of the *Bellwether moving window* was done automatically. Automating a process involves series of assumptions made which can result in biasness. This is because, assumptions do not always follow the reality on the grounds. Nevertheless, the assumptions made in this study are based on prior knowledge from previous studies [15] [9] [17] [18] in building data mining models and hence results can be trusted.

We considered three learners (SEE models) namely the Automatically Transformed Linear Model (ATLM) [1], multiple linear regression (MLR) model specifically the least squares regression [7] [6], and a Deep learning model specifically Deep Neural Networks [33]. We chose these learners because, they have been shown to improve the relative prediction accuracy.

In this study, we considered three main evaluation measures namely Mean Absolute Error (MAE), Mean Balanced Relative Error (MBRE) and Mean Inverted Balanced Relative Error (MIBRE). Even though other evaluation measures could have been used for evaluation assessment, these three measures have been proven reliable and effective in previous studies [41] [13]. Other evaluation measures such as Mean Magnitude of Relative Error (MMRE), Median Magnitude of Relative Error (MdMRE) and *PRED(k)* are misleading evaluation measures which result in biasness in model assessment as reported by Foss et al. [41].

## 9. CONCLUSION AND FUTURE WORKS

Previous studies have shown the relative effectiveness of using relevant and recently completed projects (*moving window*) from a pool of historical projects in building predictive models. Exemplary projects (*Bellwether*) representing the training set have been empirically validated to significantly improve the prediction accuracy in the domain *of* software defect prediction. Investigations on the ISBSG dataset (release 10) and the Kitchenham dataset confirm the existence of *Bellwether effect* in chronological datasets and its respective *Bellwether moving window* can be used as the training set for software effort estimation purposes. In order to achieve the best selection of relevant projects, we first provide six postulations based on empirical evidence. Secondly, we provide a step-by-step *Bellwether method* incorporating a sort and cluster approach together with Markov chains to aid in selecting the *Bellwether moving window* with defined window size (number of projects) and window age (elapsed time of projects). Results have shown that, the *Bellwether moving window* is not rare in chronological datasets and should have a window age of 2 to 2.5 calendar years and a window size of relatively large size for a better prediction accuracy. It should be noted that, the window age can vary depending on the units of the data at hand. Incorporating weighting functions on the *Bellwether moving window* affect the prediction accuracy when the window age and window size of the *Bellwether moving window* change. We observed that, the Gaussian weighting function was more advantageous on relatively large window size of 257 projects (23.4%) in the ISBSG dataset and 87 projects (61.3%) in the Kitchenham dataset.

In our future work, we intend to investigate the feasibility of using the Bellwether moving window concept in other software engineering fields for the training and validation needs of predictive models. Furthermore, we intend to extend this study to find the Bellwether moving window with defined window size and window age for other datasets more especially the Finnish dataset and other industrial projects.

### ACKNOWLEDGMENT
This work is supported in part by the General Research Fund of the Research Grants Council of Hong Kong (No. 125113,11200015 and 11214116), and the research funds of City University of Hong Kong (No. 7004683 and 7004474).

### REFERENCES

[1] P. A. Whigham, C. A. Owen, and S. G. Macdonell, "A Baseline Model for Software Effort Estimation," ACM Trans. Softw. Eng. Methodol., vol. 24, no. 3, pp. 1–11, 2015.
[2] C. Mair, M. Shepperd, and M. Jørgensen, "An analysis of data sets used to train and validate cost prediction systems," ACM SIGSOFT Softw. Eng. Notes, vol. 30, no. 4, p. 1, 2005.
[3] B. Kitchenham, B. Kitchenham, E. Mendes, and E. Mendes, "Why comparative effort prediction studies may be invalid," Proc. 5th Int. Conf. Predict. Model. Softw. Eng., no. 1, pp. 1–5, 2009.
[4] C. Lokan and E. Mendes, "Applying moving windows to software effort estimation," 2009 3rd Int. Symp. Empir. Softw. Eng. Meas. ESEM 2009, pp. 111–122, 2009.
[5] C. Lokan and E. Mendes, "Investigating the Use of Duration- Based Moving Windows to Improve Software Effort Prediction," 2012 19th Asia-Pacific Softw. Eng. Conf., pp. 818–827, 2012.
[6] S. Amasaki and C. Lokan, "A replication study on the effects of weighted moving windows for software effort estimation," Proc. 20th Int. Conf. Eval. Assess. Softw. Eng. - EASE '16, pp. 1–9, 2016.
[7] S. Amasaki and C. Lokan, "On the effectiveness of weighted moving windows: Experiment on linear regression based software effort estimation," J. Softw. Evol. Process, vol. 27, no. 7, pp. 488– 507, 2015.
[8] C. Lokan and E. Mendes, "Investigating the use of moving windows to improve software effort prediction: a replicated study," Inf. Softw. Technol., vol. 56, pp. 1063–1075, 2014.
[9] R. Krishna, T. Menzies, and W. Fu, "Too much automation? the bellwether effect and its implications for transfer learning," Proc. 31st IEEE/ACM Int. Conf. Autom. Softw. Eng. - ASE 2016, pp. 122–131, 2016.
[10] B.-C. Chen, R. Ramakrishnan, J. W. Shavlik, and P. Tamma, "Bellwether analysis," ACM Trans. Knowl. Discov. Data, vol. 3, no. 1, pp. 1–49, 2009.
[11] B.-C. Chen, R. Ramakrishnan, J. W. Shavlik, and P. Tamma, "Bellwether analysis: predicting global aggregates from local regions," Proc. 32nd Int. Conf. Very Large Data Bases, pp. 655– 666, 2006.
[12] S. J. Pan and Q. Yang, "a Survey on Transfer Learning," IEEE Trans. Knowl. Data Eng., vol. 22, no. 10, pp. 1345–1359, 2010.

[13] E. Kocaguneli, T. Menzies, and E. Mendes, "Transfer learning in effort estimation," Empir. Softw. Eng., vol. 20, no. 3, pp. 813–843, 2015.
[14] J. Huang, Y.-F. Li, and M. Xie, "An empirical analysis of data preprocessing for machine learning-based software cost estimation," Inf. Softw. Technol., vol. 67, no. June, pp. 108–127, 2015.
[15] R. P. Dobrow, "Introduction to Stochastic Processes With R," Wiley Online Libr., pp. 181–222, 2016.
[16] C. Lokan and E. Mendes, "Investigating the use of moving windows to improve software effort prediction: a replicated study," Empir. Softw. Eng., vol. 22, no. 2, pp. 716–767, 2016.
[17] B. Zhou, H. Okamura, and T. Dohi, "Enhancing performance of random testing through Markov chain Monte Carlo methods," he IEEE Trans. Comput., vol. 62, no. 1, pp. 186–192, 2013.
[18] Z. Tu and Z. Song-Chun, "Image segmentation by data-driven markov chain monte carlo," IEEE Trans. Pattern Anal. Mach. Intell., vol. 24, no. 5, pp. 657–673, 2002.
[19] H. Fischer, A history of the central limit theorem: From classical to modern probability theory. Springer Science & Business Media, 2010.
[20] K. Yao and J. Gao, "Law of Large Numbers for Uncertain Random Variables," vol. 24, no. 3, pp. 1–12, 2016.
[21] J. M. Lachin, "Introduction to sample size determination and power analysis for clinical trials," Control. Clin. Trials, vol. 2, no. 2, pp. 93–113, 1981.
[22] B. Kitchenham, "Robust statistical methods," Proc. 19th Int. Conf. Eval. Assess. Softw. Eng. - EASE '15, pp. 1–6, 2015.
[23] Y. S. Seo and D. H. Bae, On the value of outlier elimination on software effort estimation research, vol. 18, no. 4. 2013.
[24] R. Shanmugam and R. Chattamvelli, "Skewness and Kurtosis," in Statistics for Scientists and Engineers, Wiley Online Library, 2016, pp. 89–110.
[25] K. A. Brownlee and K. A. Brownlee, Statistical theory and methodology in science and engineering, Vol. 150. New York: Wiley, 1965.
[26] D. C. Montgomery, G. C. Runger, and N. F. Hubele, Engineering statistics. John Wiley & Sons, 2009.
[27] D. Pelleg and A. W. Moore, "X-means: Extending K-means with efficient estimation of the number of clusters," Proc. Seventeenth Int. Conf. Mach. Learn. table contents, pp. 727–734, 2000.
[28] T. Foss, E. Stensrud, B. Kitchenham, I. C. Society, and I. Myrtveit, "A Simulation Study of the Model Evaluation Criterion MMRE," vol. 29, no. 11, pp. 985–995, 2003.
[29] B. Kitchenham, S. L. Pfleeger, B. McColl, and S. Eagan, "An empirical study of maintenance and development estimation accuracy," J. Syst. Softw., vol. 64, no. 1, pp. 57–77, 2002.
[30] C. Lokan, T. Wright, P. R. Hill, and M. Stringer, "Organizational benchmarking using the ISBSG data repository," IEEE Softw., vol. 18, no. 5, pp. 26–32, 2001.
[31] R. D. Cook, "Detection of Influential Observation in Linear Regression," Technometrics, vol. 19, no. 1, p. 15, 1977.
[32] S. Amasaki and C. Lokan, "The evaluation of weighted moving windows for software effort estimation," in International Conference on Product Focused Software Process Improvement, 2013, vol. 7983 LNCS, pp. 214–228.
[33] S. Mensah, J. Keung, K. E. Bennin, and M. F. Bosu, "Multi-Objective Optimization for Software Testing Effort Estimation," in The 28th International Conference on Software Engineering and Knowledge Engineering, 2016, pp. 527–530.
[34] J. Schmidhuber, "Deep Learning in neural networks: An overview," Neural Networks, vol. 61, pp. 85–117, 2015.
[35] L. L. Minku and X. Yao, "Software effort estimation as a multiobjective learning problem," ACM Trans. Softw. Eng. Methodol., vol. 22, no. 4, pp. 1–32, 2013.
[36] S. Wang, T. Liu, and L. Tan, "Automatically Learning Semantic Features for Defect Prediction," Proc. 38th Int. Conf. Softw. Eng., pp. 297–308, 2016.
[37] N. Zhang and D. Shetty, "An effective LS-SVM-based approach for surface roughness prediction in machined surfaces," Neurocomputing, no. 189, pp. 35–39, 2016.
[38] F. Sarro, A. Petrozziello, and M. Harman, "Multi-objective software effort estimation," Proc. 38th Int. Conf. Softw. Eng. - ICSE '16, pp. 619–630, 2016.
[39] T. W. MacFarland and J. M. Yates, "Kruskal–Wallis H-Test for Oneway Analysis of Variance (ANOVA) by Ranks," in Introduction to Nonparametric Statistics for the Biological Sciences Using R, Springer International Publishing, 2016, pp. 177–211.
[40] M. Shepperd and S. MacDonell, "Evaluating prediction systems in software project estimation," Inf. Softw. Technol., vol. 54, no. 8, pp. 820–827, 2012.
[41] T. Foss, E. Stensrud, B. Kitchenham, and I. Myrtveit, "A simulation study of the model evaluation criterion MMRE," Softw. Eng. IEEE Trans., vol. 29, no. 11, pp. 985–995, 2003.